\documentclass[prb,twocolumn,showpacs,preprintnumbers,amsmath,amssymb]{revtex4}
\usepackage{graphicx}
\usepackage{bm}

\begin{document}

\preprint{ }

\title{Rectified dc voltage versus
magnetic field in a superconducting asymmetric
figure-of-eight-shaped microstructure}


\author{V.~I. Kuznetsov}
\email{kvi@ipmt-hpm.ac.ru}
\author{A.~A. Firsov}
\author{S.~V. Dubonos}
\affiliation{Institute of Microelectronics Technology and High
Purity Materials, Russian Academy of Sciences, Chernogolovka,
Moscow Region 142432, Russia}

\date{\today}

\begin{abstract}
We have measured periodic oscillations of rectified dc voltage
versus magnetic field $V_{dc}(B)$ in a superconducting aluminum
thin-film circular-asymmetric figure-of-eight microstructure
threaded by a magnetic flux and biased with a sinusoidal
alternating current (without a dc component) near the critical
temperature. The Fourier spectra of these $V_{dc}(B)$ functions
contain fundamental frequencies representing periodic responses of
the larger and smaller asymmetric circular loops, composing the
microstructure, to the magnetic field. The higher harmonics of the
obtained fundamental frequencies result from the non-sinusoidal
character of loop circulating currents. The presence of the
difference and summation frequencies in these spectra points to
the interaction between the quantum states of both loops.
Magnitudes of the loop responses to the bias ac and magnetic field
vary with temperature and the bias current amplitude, both in
absolute values and with respect to each other. The strongest loop
response appears when the average resistive state of the loop
corresponds to the midpoint of the superconducting-normal phase
transition.
\end{abstract}


\maketitle

\section{INTRODUCTION}
Superconducting loops interrupted by tunnel junctions are used in
superconducting quantum interference devices \cite{sit31}and
typical superconducting flux qubits. \cite{sit10} This work deals
with double superconducting circular-asymmetric loops without
tunnel contacts. For the first time, it has been found \cite{sit1}
that a single superconducting asymmetric circular loop is a simple
and very efficient rectifier of ac voltage.

Nonzero time-averaged rectified dc voltage $V_{dc}(B)$ was
experimentally observed \cite{sit1} in a single superconducting
aluminum asymmetric circular loop threaded by a magnetic flux
$\Phi$ and biased with a sinusoidal alternating current (without a
dc component) with an amplitude close to critical and frequencies
up to 1 MHz at temperatures slightly below the superconducting
critical temperature $T_{c}$. This $V_{dc}(B)$ voltage as a
function of magnetic field $B$ oscillates with the period $\Delta
B=\Phi_{0}/S$, where $\Phi_{0}$ is the superconducting magnetic
flux quantum and $S$ is the effective loop area. \cite{sit2,pit1}
It was also shown \cite{sit1} that the magnitude of rectified
voltage could easily be increased by a serial connection of such
loops.

The $V_{dc}(B)$ oscillations as well as the Little-Parks (LP) ones
\cite{sit3} are determined by the requirement of superconducting
fluxoid quantization, \cite{sit24} or, set it another way,
quantization of the circulation of a total quantum angular
momentum along a closed contour. Unlike $R(B)$ oscillations in the
LP effect, \cite{sit3} $V_{dc}(B)$ voltage is the odd function of
magnetic field. \cite{sit1}

The work by S. Dubonos {\it et al.} \cite{sit1} supplies indirect
experimental arguments that $V_{dc}(B)$ voltage in a single
asymmetric circular loop is directly proportional to a
magnetic-field-dependent circulating current  $I_{R}(B)$ of the
loop. If this is the case, the measurement of $V_{dc}(B)$ (as we
think) permits a complete determination of the quantum state of
the loop, viz., both the magnitude and direction of the
circulating current at different values of the magnetic field.

It would be of interest to investigate the quantum behavior of
more complicated asymmetric multiply connected \ structures by
measuring rectified dc voltage $V_{dc}(B)$ as a function of the
magnetic field. The $V_{dc}(B)$ measurements could be expected to
allow the quantum state determination of each loop and possible
interaction between the loops in a system of serial
circular-asymmetrical loops with different areas. Therefore, we
measured the $V_{dc}(B)$ voltage both in a system of two serial
loops coupled by a wire (results obtained will be presented
elsewhere) and in a system of two directly coupled loops forming a
figure-of-eight structure (Fig. \ref{image}).

This work was inspired by a supposition made in Ref.
\onlinecite{sit32}. The subject discussed was the possibility to
use an asymmetric circular loop like the one considered in Ref.
\onlinecite{sit1}, although with extremely thin walls, as an
element for a flux qubit without tunnel contacts. At present, we
do not know of any experimental work dealing with a
superconducting flux qubit without tunnel contacts.

Quantum \ phase-slip centers (QPSCs) \cite{sit14,sit15,sit16} can
be formed in nanostructures with cross-sectional dimensions of
less than $10$ nm and at temperatures below $0.5T_{c}$. Another
superconducting flux qubit with QPSCs instead of tunnel contacts
has recently been theoretically considered. \cite{sit17} We
believe that an asymmetric figure-of-eight structure, like the one
considered here (Fig. \ref{image}), but with extremely thin loop
walls, can be a prototype of two directly coupled different flux
qubits. Needless to say that quantum tunneling between two
distinct macroscopic quantum states cannot be realized in a
structure with the geometry and external parameters used here.

Quantum behavior of systems of two superconducting loops as a
function of the magnetic field was studied earlier. A double
superconducting loop composed of two equal squares having a common
side was used to study the features of the $T_{c}(B)$ function.
\cite{sit7,sit8} Magnetic coupling between two superconducting
coaxial square loops was experimentally studied. \cite{sit9} This
work essentially differs from those reported in Refs.
\onlinecite{sit7,sit8,sit9} by a circular-asymmetric geometry of
the structure (Fig. \ref{image}). Unlike symmetric structures,
\cite{sit7,sit8,sit9} the geometry of the structure provides a
chance to use $V_{dc}(B)$ measurements to detect quantum states of
both a figure-of-eight double loop taken as a whole and each
circular loop individually.

The goal of the work is to experimentally study  the quantum
behavior of rectified dc voltage $V_{dc}(B)$ versus perpendicular
magnetic field and bias sinusoidal low-frequency current (without
a dc component) at temperatures slightly below $T_{c}$ in a
superconducting aluminum figure-of-eight structure (Fig.
\ref{image}). Moreover, we hope to evaluate relative contributions
of both circular loops of the structure into the total dc voltage
and to detect a presupposed interaction in the structure.

\section{SAMPLES AND EXPERIMENTAL PROCEDURE}
\begin{figure}
\includegraphics[width=1.0\columnwidth]{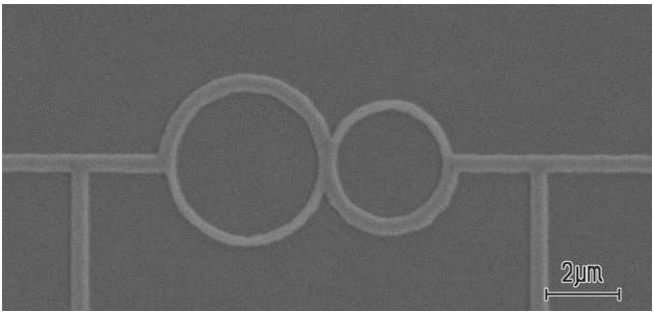}
\caption{\label{image} Scanning electron microscopy image of the
structure. The scale bar: $2$ $\mu$m.}
\end{figure}

Structures were fabricated by thermal aluminum deposition onto Si
substrates using the lift-off process of electron-beam
lithography. The NANOMAKER program package with correction for the
proximity effect was employed. The central region of the structure
(Fig. \ref{image}) is figure-of-eight shaped and circular
asymmetric, with the widths of wide wires $w_{w}=0.47$ $\mu$m and
narrow wires $w_{n}=0.24$ $\mu$m and the film thickness $d=70$ nm.
The structure consists of two circularly asymmetric loops of
different areas having a common area. The average area of the
larger loop determined as a sum of average areas of the upper and
lower semiloops is equal to $S^{g}_{L}=14.51$ $\mu\rm m^{2}$. The
area of the smaller loop is $S^{g}_{S}=8.51$ $\mu\rm m^2$. The
effective mean radii of the larger and smaller loops calculated
from $S^{g}_{L}$ and $S^{g}_{S}$ are equal to $r_{L}=2.15$ $\mu$m
and $r_{S}=1.65$ $\mu$m, respectively.

The total normal-state resistance at $T=4.2$ K is $R_{4.2}=8.39$
$\Omega$. The ratio of room-temperature resistance to the helium
one is $R_{300}/R_{4.2}=2.22$, and the sheet resistance is
$R_{\square}=0.33$ $\Omega$, hence, the resistivity is
$\rho=2.37\times 10^{-8}$ $\Omega$ m. From the known mean value of
the product \cite{sit21} $\rho l=6\times 10^{-16}$ $\Omega$ $\rm
m^2$, the electron mean free path in the structure is $l=25$ nm.
The superconducting coherence length of pure aluminum at zero
temperature is $\xi_{0}=1.6$ $\mu$m. Because $l \ll \xi_{0}$ in
our structure, it can be regarded as a "dirty" superconductor.
\cite{sit22} The critical superconducting temperature
$T_{c}=1.324\pm 0.001$ K was determined in the midpoint of
normal-superconducting transition $R(T)$ at very small currents in
a zero field.

At temperatures slightly below $T_{c}$, the temperature-dependent
coherence length of the dirty superconductor and the field
penetration depth are determined by the expressions:
\cite{sit22,sit23} $\xi(T)=\xi(0)(1-T/T_{c})^{-1/2}$ and
$\lambda_{d}(T)=\lambda_{d}(0)(1-T/T_{c})^{-1/2}$, respectively,
where $\xi(0)=0.85(\xi_{0}l)^{1/2}$ and
$\lambda_{d}(0)=0.615\lambda_{L}(0)(\xi_{0}/l)^{1/2}$. Here,
$\lambda_{L}(0)$ is the London penetration depth of a pure
superconductor at zero temperature. In pure aluminum
superconductors, \cite{sit23} $\lambda_{L}(0)=0.05$ $\mu$m. Hence,
$\xi(0)=0.17$ $\mu$m and $\lambda_{d}(0)=0.25$ $\mu$m in our
structure.

\begin{figure*}\includegraphics[width=3.35 in]{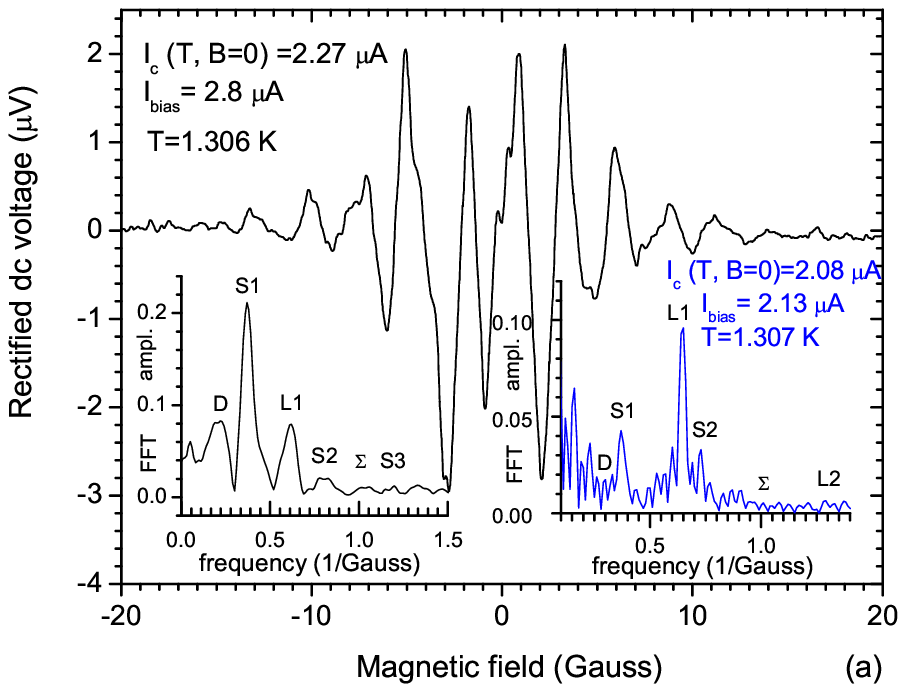}\includegraphics[width=3.35 in]{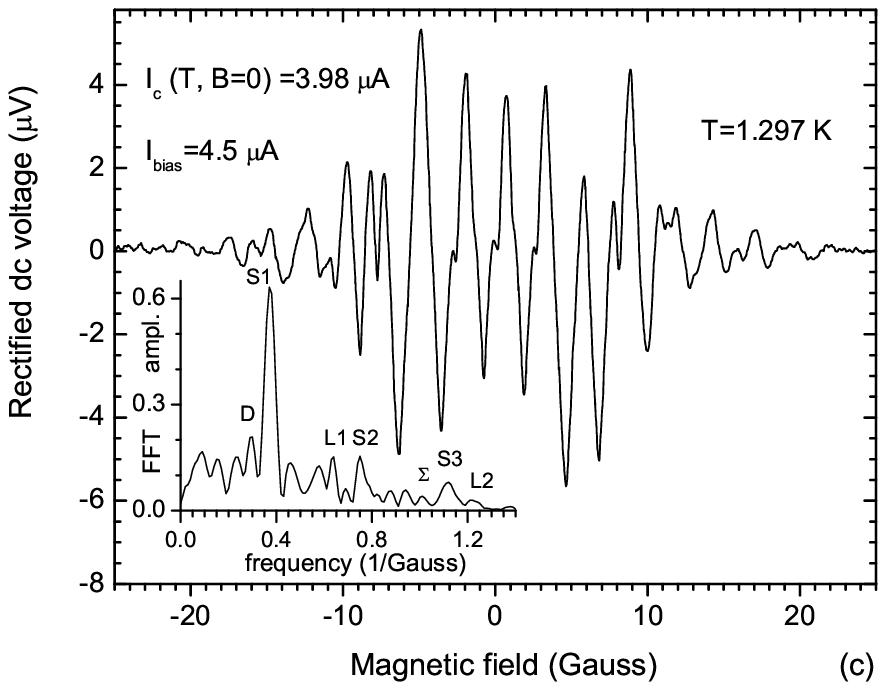}
\includegraphics[width=3.35 in]{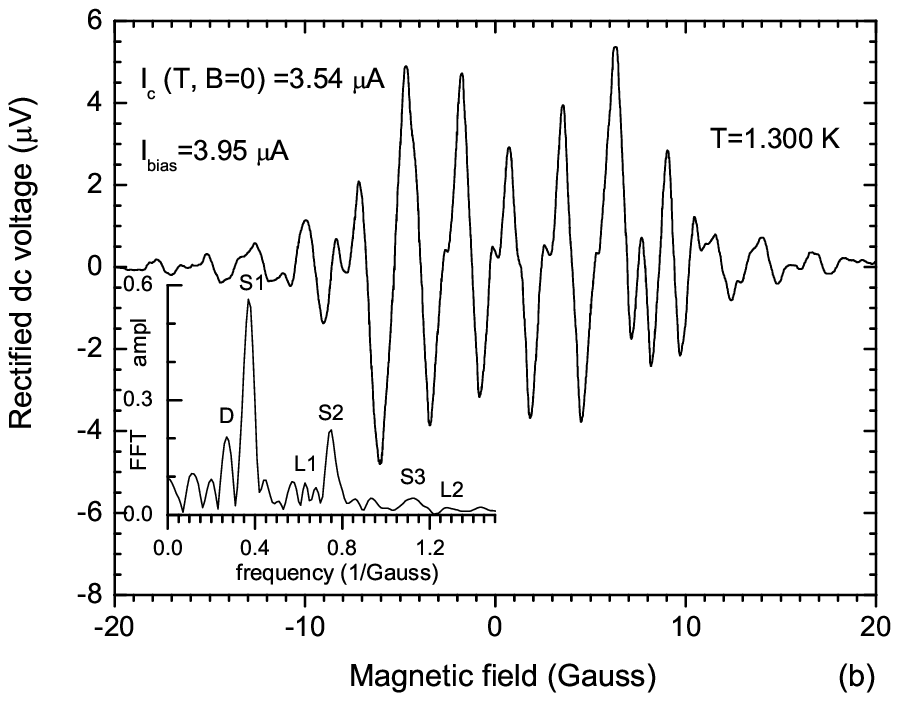}\includegraphics[width=3.35 in]{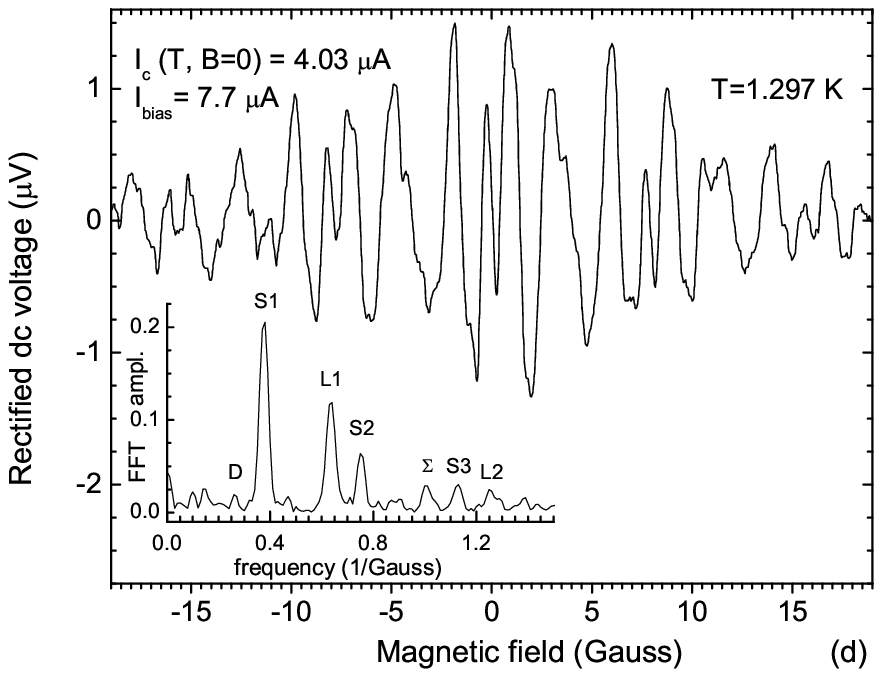}
\caption{\label{voltage} (Color online) [(a)-(d)] Experimental
$V_{dc}(B)$ plots. The left insets show the FFT spectra of the
$V_{dc}(B)$ curves. The right inset of (a) presents the FFT
spectrum of an experimental $V_{dc}(B)$ curve measured at
parameters given in the inset [this $V_{dc}(B)$ curve is not shown
here]. The symbols $S1$, $S2$, $S3$, and $L1$, $L2$ in the insets
refer to the spectral peaks corresponding to the fundamental
frequencies and their higher harmonics for the smaller and larger
loops, respectively. The symbols $D$ and $\Sigma$ refer to the
spectral peaks corresponding to the difference and summation
frequencies, respectively.}
\end{figure*}

We present four-probe measurements of the $V_{dc}(B)$ oscillations
in the structure (Fig. \ref{image}). Magnetic field $B$ was
perpendicularly applied to the structure surface. The structure
was periodically switched to the resistive state by a sinusoidal
current (without a dc component) $I_{bias}(t)=I_{bias}\sin(2\pi\nu
t)$, with the amplitude $I_{bias}$ close to critical at
frequencies $\nu$ of up to $1$ MHz at temperatures slightly below
$T_{c}$. The $V_{dc}(B)$ voltage was measured at slowly varying
$B$ with a sweep period $\Delta t_{B}$. The $V_{dc}(B)$ was equal
to the time-averaged momentary pulsating voltage $V(t)$ over a
large time interval $\Delta t_{L}$. Moreover, the condition
$\Delta t_{B}>20\Delta t_{L}>400\Delta t_{I}$ was valid in all the
experimental cases. Here, $\Delta t_{I}$ is the period of bias ac.
So, the measured voltage $V_{dc}(B)$ was practically equal to
$\frac{1}{\Delta t_{I}}\int^{\Delta t_{I}}_{0}V(t)dt$. The
experimental $V_{dc}(B)$ function is probably the result of
multiple time-averaged measurements of the structure quantum
state. Similar $V_{dc}(B)$ functions were obtained for six
structures of similar geometries and similar external parameters.

\section{RESULTS AND DISCUSSION}
Figure \ref{voltage} shows the $V_{dc}(B)$ curves at different
temperatures slightly below $T_{c}$ in the structure biased with a
sinusoidal current of an amplitude $I_{bias}$, close to critical,
and a frequency of $1.5$ kHz (without a dc component). The
magnitudes of critical bias current $I_{c}(T,B=0)$ are shown in
Fig. \ref{voltage} at several temperatures $T$ in the zero field.
In higher fields, the $V_{dc}(B)$ oscillations were damped due to
the suppression of the superconducting order parameter in a wire
of finite width. For a detailed analysis, fast Fourier transforms
(FFTs) of the $V_{dc}(B)$ functions were calculated. The left
insets of Fig. \ref{voltage} show the FFT spectra of the
$V_{dc}(B)$ functions. The right inset of Fig. \ref{voltage}(a)
presents the FFT spectrum of $V_{dc}(B)$ curve (the curve is not
shown) with external parameters given in the inset. FFT spectra
were obtained using $two^{12}$ uniformly distributed points in
magnetic fields ranging from $-50$ to $+50$ G. In the insets,
numerous peaks are observed at certain frequencies which are the
inverse magnitudes of different oscillation periods of the
$V_{dc}(B)$ functions. The values of fundamental frequencies
(first harmonics) $f_{S1}$ and $f_{L1}$ are equal to the inverse
magnitudes of the oscillation periods corresponding to the
effective areas of the smaller $S_{S}$ and the larger $S_{L}$
loops of the figure-of-eight structure and are
\begin{equation}
f_{S1}=1/\Delta B_{S}=S_{S}/\Phi_{0}~,~~\rm ~~f_{L1}=1/\Delta
B_{L}=S_{L}/\Phi_{0}~,
\end{equation} respectively.

Using average geometric values of areas for the smaller
$S^{g}_{S}$ and the larger $S^{g}_{L}$ circular loops instead of
the values of effective areas $S_{S}$ and $S_{L}$, we obtain
"geometric" values of the fundamental frequencies
$f^{g}_{S1}=0.41$ ${\rm G}^{-1}$ and $f^{g}_{L1}=0.70$ ${\rm
G}^{-1}$. Indeed, the FFT spectra exhibit corresponding peaks at
frequencies of $0.37$ and $0.63$ ${\rm G}^{-1}$ close to these
geometric values. The measured values of the fundamental Fourier
frequencies, $f_{S1}$ and $f_{L1}$, are smaller than their
geometric values. The difference is probably due to the fact that
the effective area of the loop is smaller than the averaged
geometric one.

We found that quantum resistive contributions of both loops into
the total rectified voltage can vary with temperature and bias
current amplitude, both in absolute values and with respect to
each other (Fig. \ref{voltage}). Unexpectedly, the amplitude of
the peak corresponding to the fundamental frequency of the larger
loop turned out to considerably exceed the amplitude of the peak
corresponding to the fundamental frequency of the smaller loop
[the right inset of Fig. \ref{voltage}(a)]. In addition to the
fundamental frequencies $f_{S1}$ and $f_{L1}$, the Fourier spectra
contain higher harmonics of these frequencies, $f_{Sm}=mf_{S1}$
and $f_{Lm}=mf_{L1}$, where $m=2,3,4...$, and difference and
summation frequencies, $f_{D}=f_{L1}-f_{S1}$ and
$f_{\Sigma}=f_{L1}+f_{S1}$, respectively.

Let us now discuss the results obtained. It can be expected that
at temperatures slightly below $T_{c}$, the most efficient
rectification of alternating voltage is realized when a joint
effect of bias ac $I_{bias}(t)$ and loop circulating current
$I_{R}(B)$ periodically switch the structure from a
superconducting $S$ state to that with finite resistance (close to
normal $N$ state) and back. Moreover, a time-averaged resistive
state of the structure would correspond to the midpoint of the
$S$-$N$ transition.

We guess that there are two reasons as to why the difference
between the magnitudes of circulating currents in the loops
forming the figure-of-eight-shaped structure can result in
different relative resistive contributions of the loops into the
total rectified dc voltage in the structure. On the one hand, if
$V_{dc}(B) \propto I_{R}(B)$ in each circular loop, then the
higher magnitude of the loop circulating current should result in
a higher contribution into the total dc voltage. On the other
hand, the relative contribution of each loop to the total dc
voltage should be a nonmonotonic function of the bias ac amplitude
\cite{sit1} $I_{bias}$. Moreover, the contributions of the loops
should reach their maximum at different values of the $I_{bias}$
amplitude close to the critical current. A larger contribution of
a loop should be expected when the average resistive state of the
loop is closer to the midpoint of the $S$-$N$ transition.

To check the assumption that the difference between the magnitudes
of circulating currents results in different relative
contributions of the loops, we calculated circulating currents of
each loop composing the structure at four values of $T$ shown in
Fig. \ref{voltage} and in the right inset of Fig.
\ref{voltage}(a). We used the Ginzburg-Landau theory for
finite-wall-thickness asymmetric circular loops, neglecting radial
variations of the order parameter and self-field generated by
superconducting currents, which is reasonable because
$\lambda_{d}(T)>d=70$ nm, $\xi(T)\approx 1$ $\mu\rm m>w_{n}=0.24$
$\mu$m, and $w_{w}=0.47$ $\mu$m. The calculated circulating
currents of the smaller $I_{RS}(B)$ and the larger $I_{RL}(B)$
loops are not strongly harmonic functions. In higher fields,
$I_{RS}(B)$ and $I_{RL}(B)$ oscillations as well as the
$V_{dc}(B)$ ones were damped.

Provided that $V_{dc}(B) \propto I_{R}(B)$ in each asymmetric
circular loop, nonharmonicity of loop circulating currents should
result in the appearance of higher harmonics of loop fundamental
frequencies in the FFT spectra of the total dc voltage $V_{dc}(B)$
measured in the structure. At temperatures $T=1.307$, $1.306$,
$1.300$, and $1.297$ K, the maximum magnitudes of the calculated
circulating currents were equal to $\vert I_{RS}\vert = 0.14$,
$0.22$, $0.66$, $0.83$, and $\vert I_{RL}\vert =0.47$, $0.55$,
$1.06$, $1.25$ in the smaller and the larger loops, respectively.
Here, the current values are given in microampere.

So at these temperatures, a higher circulating current corresponds
to a larger diameter of the loop. On the contrary, for circular
loops (cylinders) with infinitely thin walls, \cite{sit24} the
smaller the structure diameter, the higher the circulating current
is. In our case, both walls of each asymmetric loop have two
finite thicknesses, $w_{n}$ and $w_{w}$. Therefore, nonzero terms
containing $w/2R$ are included in the expression for circulating
current. \cite{sit2} A larger value of $w/2R$ can result in a
smaller value of the circulating current.

Using the maximum values of the calculated circulating currents
and experimental magnitudes of the critical current
$I_{c}(T,B=0)$, we can estimate how close the average resistive
state of each loop with all parameters shown in Fig. \ref{voltage}
and the right inset of Fig. \ref{voltage}(a) can be to the
midpoint of the $S$-$N$ transition. Then, presupposed relative
contributions of both loops to the total rectified dc voltage can
be evaluated. The estimations showed that for each average
resistive state of the loop there is a certain point in the
$S$-$N$ transition.

At all parameters [except the parameters for the right inset of
Fig. \ref{voltage}(a)], the point in the $S$-$N$ transition
corresponding to the average resistive state of the smaller loop
is nearer to the midpoint of the $S$-$N$ transition than the point
corresponding to the average resistive state of the larger loop.
Therefore, the contribution from the smaller loop can be expected
to be higher than that from the larger one. With the parameters
shown in the right inset of Fig. \ref{voltage}(a), the point
corresponding to the average resistive state of the larger loop is
closer to the midpoint of the $S$-$N$ transition, whereas the
point corresponding to the average resistive state of the smaller
loop is closer to the region of the superconducting state. Then,
the contribution of the larger loop should be much greater than
that of the smaller one. With the parameters given in Fig.
\ref{voltage}(d), the bias ac amplitude $I_{bias}$ considerably
exceeds the critical current $I_{c}(T,B=0)$, and the average
resistive states of both loops are very close to the normal state.
This results in both a radical decrease in absolute values of both
loop contributions to the total rectified voltage and a decrease
in the difference between relative contributions of both loops.
Indeed, these estimated relative contributions of the loops
forming the structure to the total rectified dc voltage agree with
experimental relative contributions [Fig. \ref{voltage} and the
right inset of Fig. \ref{voltage}(a)].

The FFT spectra of the $V_{dc}(B)$ functions contain frequencies
close to the summation  $f_{\Sigma}$ and difference $f_{D}$
frequencies. Because $f_{\Sigma}=1/\Delta
B_{\Sigma}=f_{S1}+f_{L1}=(S_{S}+S_{L})/\Phi_{0}$ and
$f_{D}=1/\Delta B_{D}=f_{S1}-f_{L1}=(S_{L}-S_{S})/\Phi_{0}$, the
$f_{\Sigma}$ and $f_{D}$ frequencies are directly proportional to
the sum and difference of the loop effective areas, respectively.
The presence of the $f_{\Sigma}$ and $f_{D}$ frequencies in the
spectra points out to the interaction (nonlinear coupling) between
the loops.

Let us consider possible mechanisms of the interaction. The
magnetic inductive coupling between both circular loops of the
figure-of-eight structure can qualitatively explain the appearance
of the $f_{D}$ frequency. However, the interaction should be weak
because of the structure geometry. So, the inductive coupling
between the two loops composing the figure-of-eight structure is
ten times weaker than that between coaxial loops of the same
dimensions.

Apart from circulating currents of both circular loops, an
additional periodic magnetic-field-dependent closed current can
appear along a figure-of-eight contour for which the requirement
of superconducting fluxoid quantization is also valid. This
additional closed current can qualitatively explain the appearance
of the  $f_{\Sigma}$ frequency. The magnitude of the additional
current should be small. Indeed, the magnitude of the FFT spectral
peak corresponding to the $f_{\Sigma}$ frequency was lower than
the peak magnitude corresponding to the $f_{D}$ frequency.

Another possible reason is that the difference and summation
frequencies can be due to electrodynamic interaction between the
circular loops, realized through a common bias ac. A
characteristic longitudinal scale of the interaction should be
expected to equal the penetration depth of a nonuniform electric
field \cite{sit23} into the superconducting structure. The scale
considerably exceeds the superconducting coherence length
$\xi(T)$. Together with the oscillations of the loop circulating
currents, weak oscillations of time-averaged absolute value of the
order parameter versus magnetic field occur in both loops. When
the bias ac with an amplitude close to critical passes through the
figure-of-eight structure, both superconducting and normal
components of the current become doubly modulated with the
fundamental frequencies corresponding to both circular loops.
Because of electrodynamic coupling between the bias ac and both
loop circulating currents, difference, summation, and other
combination frequencies can arise. To better understand the
mechanisms of this interaction, further experiments will be
carried out.

\section{CONCLUSION}
In conclusion, quantum oscillations of a rectified dc voltage
$V_{dc}(B)$ as a function of magnetic field were measured in a
superconducting circular-asymmetric figure-of-eight structure. The
Fourier analysis of the $V_{dc}(B)$ oscillations revealed relative
contributions of both loops, forming the structure, into the total
dc voltage. These contributions varied with the bias ac and
temperature both in absolute magnitude and with respect to each
other. The contribution of the loop is maximum when the average
resistive state of the loop corresponds to the midpoint of the
$S$-$N$ phase transition. An interaction between quantum states
corresponding to the two circular loops was found. Magnetic
coupling, formation of an additional figure-of-eight contour for a
periodic magnetic-field-dependent closed current, and
electrodynamic coupling through a common bias ac can be the
possible mechanisms of the interaction between the loops.

\section{ACKNOWLEDGMENTS}
The authors are grateful to V.~Tulin, A.~Nikulov, V.~Gurtovoi,
M.~Chukalina, M.~Skvortsov, A.~Alexandrov, Ya.~Greenberg,
E.~Il'ichev, and V.~Moshchalkov for helpful discussions, and
P.~Shabelnikova and A.~Chernih for technical help. The work was
financially supported in the framework of the program
"Computations based on novel physical quantum algorithms,"
Information Technologies and Computer Systems Department of the
Russian Academy of Sciences.


\begin{thebibliography}{99}

\bibitem{sit31}

A. Barone, G. Paterno, {\it Physics and Applications of the
Josephson Effect} (Willey-Interscience, New York, 1982).

\bibitem{sit10}

J.~E. Mooij, T.~P. Orlando, L. Levitov, L. Tian, C.~H.
van~der~Wal, and S. Lloyd, Science {\bf285}, 1036 (1999).

\bibitem{sit1}

S.~V. Dubonos, V.~I. Kuznetsov, I.~N. Zhilyaev, A.~V. Nikulov, and
A.~A. Firsov, JETP Lett. {\bf77}, 371 (2003).

\bibitem{sit2}

R.~P. Groff and R.~D. Parks, Phys. Rev. {\bf176}, 567 (1968).

\bibitem{pit1}

R.~M. Arutyunyan and G.~F. Zharkov, J. Low Temp. Phys.
\textbf{52}, 409 (1983).

\bibitem{sit3}

W.~A. Little and R.~D. Parks, Phys. Rev. Lett. {\bf9}, 9 (1962).

\bibitem{sit24}

M. Tinkham, {\it Introduction to Superconductivity} (McGraw-Hill,
New York, 1975).

\bibitem{sit32}

V.~I. Kuznetsov and V.~A. Tulin, Proceedings of the First
International Conference on Fundamental Problems of HTS in Russia,
Zvenigorod, Moscow 2004 (unpublished), Sec. ~A, p. ~305.

\bibitem{sit14}

A.~D. Zaikin, D.~S. Golubev, A. van~Otterlo, and G.~T. Zimanyi,
Phys. Rev. Lett. {\bf78}, 1552 (1997).

\bibitem{sit15}

A. Bezryadin, C.~N. Lau, and M. Tinkham, Nature (London) {\bf404},
971 (2000).

\bibitem{sit16}

C.~N. Lau, N. Markovic, M. Bockrath, A. Bezryadin, and M. Tinkham,
Phys. Rev. Lett. {\bf87}, 217003 (2001).

\bibitem{sit17}

J.~E. Mooij and C.~J.~P.~M. Harmans, New J. Phys. {\bf7}, 219
(2005).

\bibitem{sit7}

V. Bruyndoncx, \ C.~Strunk, V.~V. Moshchalkov, C. Van~Haesendonck,
and Y. Bruynseraede, Europhys. Lett. {\bf36}, 449 (1996).

\bibitem{sit8}

V.~M. Fomin, J.~T. Devreese, V. Bruyndoncx, and V.~V. Moshchalkov,
Phys. Rev. B {\bf62}, 9186 (2000).

\bibitem{sit9}

M. Morelle, V. Bruyndoncx, R. Jonckheere, and V.~V. Moshchalkov,
Phys. Rev. B {\bf64}, 064516 (2001).

\bibitem{sit21}

K.~Yu. Arutyunov, D.~A. Presnov, S.~V. Lotkhov, A.~B. Pavolotski,
and L. Rinderer, Phys. Rev. B {\bf59}, 6487 (1999).

\bibitem{sit22}

P.~G. de~Gennes, {\it Superconductivity of Metals and Alloys}
(Benjamin, New York, 1966).

\bibitem{sit23}

V.~V. Schmidt, {\it The Physics of Superconductors}, edited by P.
Muller and A.~V. Ustinov (Springer-Verlag, Berlin, 1997).

\end{thebibliography}
\end{document}